# HIRES, the high-resolution spectrograph for the ELT

Alessandro Marconi[1,2], Manuel Abreu[14,15], Vardan Adibekyan[19,31], Matteo Aliverti[6], Carlos Allende Prieto[8,29], Pedro J. Amado[9], Manuel Amate[8], Etienne Artigau[32], Sergio R. Augusto[10], Susana Barros[19,31], Santiago Becerril[9], Björn Benneke[32], Edwin Bergin[34], Philippe Berio[25], Naidu Bezawada[40], Isabelle Boisse[12], Xavier Bonfils[37], Francois Bouchy[13,12], Christopher Broeg[41], Alexandre Cabral[14,15], Rocio Calvo-Ortega[9], Bruno Leonardo Canto Martins[16], Bruno Chazelas[13], Andrea Chiavassa[25], Lise B. Christensen[20], Roberto Cirami[3], Igor Coretti[3], Stefano Cristiani[3], Vanderlei Cunha Parro[10], Guido Cupani[3], Izan de Castro Leão[16], José Renan de Medeiros[16], Marco Antonio Furlan de Souza[10], Paolo Di Marcantonio[3], Igor Di Varano[17], Valentina D'Odorico[3], René Doyon[32], Holger Drass[18,28], Pedro Figueira[30,19], Ana Belen Fragoso[8], Johan Peter Uldall Fynbo[20], Elena Gallo[34], Matteo Genoni[6], Jonay I. González Hernández[8,29], Martin Haehnelt[35], Julie Hlavacek-Larrondo[33], Ian Hughes[13], Philipp Huke[22], Andrew Humphrey[19], Hans Kjeldsen[24], Andreas Korn[26], Driss Kouach[36], Marco Landoni[6], Jochen Liske[27], Christophe Lovis[13], David Lunney[11], Roberto Maiolino[4], Lison Malo[32], Thomas Marquart[26], Carlos J. A. P. Martins[19,38], Elena Mason[3], John Monnier[34], Manuel A. Monteiro[19], Christoph Mordasini[41], Tim Morris[21], Graham J. Murray[21], Andrzej Niedzielski[23], Nelson Nunes[14,15], Ernesto Oliva[2], Livia Origlia[5], Enric Pallé[8,29], Giorgio Pariani[6], Phil Parr-Burman[11], José Peñate[8], Francesco Pepe[13], Enrico Pinna[2], Nikolai Piskunov[26], Jose Luis Rasilla[8], Phil Rees[11], Rafael Rebolo[8,29], Ansgar Reiners[22], Marco Riva[6], Sylvain Rousseau[25], Nicoletta Sanna[2], Nuno C. Santos[19,31], Mirsad Sarajlic[41], Tzu-Chiang Shen[18], Francesca Sortino[7], Danuta Sosnowska[13], Sérgio Sousa[19], Eric Stempels[26], Klaus G. Strassmeier[17], Fabio Tenegi[8], Andrea Tozzi[2], Stephane Udry[13], Luca Valenziano[7], Leonardo Vanzi[18], Michael Weber[17], Manfred Woche[17], Marco Xompero[2], Erik Zackrisson[26], María Rosa Zapatero Osorio[39]

[1]Dip. di Fisica e Astronomia, Univ. di Firenze, via G. Sansone 1, I-50019, Sesto F.no (FI), Italy
[2]INAF-Osservatorio Astrofisico di Arcetri, Largo E. Fermi 2, I-50125, Firenze, Italy
[3]INAF Osservatorio Astronomico di Trieste, Via Giambattista Tiepolo 11, 34131 - Trieste Italy
[4]Cavendish Laboratory, Univ. of Cambridge, JJ Thomson Avenue, Cambridge CB3 0HE, UK
[5]INAF-Osservatorio Astronomico di Bologna, Via Ranzani, 1, 40127, Bologna, Italy
[6]INAF-Osservatorio Astronomico di Brera, Via Bianchi 46, I-23807 Merate, Italy
[7]INAF-Osservatorio di Astrofisica e Scienze dello Spazio di Bologna, via P. Gobetti 93/3, 40129 Bologna, Italy
[8]Instituto de Astrofisica de Canarias (IAC), C/ Vìa Lactea, s/n E-38205, La Laguna, Tenerife, Spain
[9]Instituto de Astrofisica de Andalucia-CSIC Glorieta de la Astronomia s/n, 18008, Granada, Spain
[10]IMT - Instituto Mauá de Tecnologia, Praça Mauá, 1 - Mauá, São Caetano do Sul - SP Brazil, 09580-900
[11]UK Astronomy Technology Centre (part of the Science and Technology Facilities Council), Blackford Hill, Edinburgh, EH9 3HJ, UK
[12]Aix Marseille Univ, CNRS, CNES, LAM, Marseille, France
[13]Dèpartement d'Astronomie, Universitè de Geneve, Chemin des Maillettes 51, Sauverny, CH-1290 Versoix, Switzerland
[14]Instituto de Astrofísica e Ciências do Espaço, Universidade de Lisboa, Campo Grande 1749-016 Lisboa Portugal
[15]Departamento de Física, Faculdade de Ciências, Universidade de Lisboa, Campo Grande 1749-016 Lisboa Portugal
[16]Board of Observational Astronomy, Federal University of Rio Grande do Norte, Campus Universit´ario 59078-970, Natal RN, Brasil
[17]Leibniz Institute for Astrophysics Potsdam (AIP), An der Sternwarte 16, D-14482 Potsdam, Germany

[18]Centro de Astro Ingenieria, Pontificia Universidad Catolica de Chile, Avda. Libertador Bernardo O'Higgins 340 - Santiago de Chile
[19]Instituto de Astrofísica e Ciências do Espaço, Universidade do Porto, CAUP, Rua das Estrelas, PT4150-762 Porto, Portugal
[20]Cosmic Dawn Center, Niels Bohr Institute, Copenhagen University, DK-2100 Copenhagen O, Denmark
[21]Centre for Advanced Instrumentation, Department of Physics, Durham University, South Road, Durham, DH1 3LE, UK
[22]Institute for Astrophysics University of Göttingen, Friedrich-Hund-Platz 1 37077 Göttingen, Germany
[23]Institute of Astronomy, Nicolaus Copernicus University in Torun, Gagarina 11, 87-100 Torun, Poland
[24]Department of Physics and Astronomy, Ny Munkegade 120, building 1520, 527, 8000 Aarhus C, Denmark
[25]Laboratoire Lagrange, Université Côte d'Azur, Observatoire de la Côte d'Azur, CNRS, Blvd de L'Observatoire CS34229,06004 Nice Cedex 4, France
[26]Division of Astronomy and Space Physics, Department of Physics and Astronomy, Uppsala University, Box 516, S-75120 Uppsala, Sweden
[27]Hamburger Sternwarte, Universität Hamburg, Gojenbergsweg 112, D-21029 Hamburg
[28]Millennium Institute of Astrophysics, Santiago, Chile
[29]Universidad de La Laguna (ULL), Departamento de Astrofisica, E-38206 La Laguna, Tenerife, Spain
[30]European Southern Observatory, Alonso de Cordova 3107, Vitacura, Santiago, Chile
[31]Departamento de Física e Astronomia, Faculdade de Ciências, Universidade do Porto, Rua do Campo Alegre, 4169-007 Porto, Portugal
[32]Institut de Recherche sur les Exoplanètes and Observatoire du Mont-Mégantic, département de physique, Université de Montréal, CP 6128 Succ. Centre-ville, H3C 3J7, Montréal, QC, Canada
[33]Département de physique, Université de Montréal, CP 6128 Succ. Centre-ville, H3C 3J7, Montréal, QC, Canada.
[34]Department of Astronomy, University of Michigan, 323 West Hall, 1085 S. University Avenue, Ann Arbor, MI 48109, USA
[35]Kavli Institute for Cosmology and Institute of Astronomy, Madingley Road, Cambridge CB3 0HA, UK
[36]CNRS, OMP, Université de Toulouse, 14 Avenue Belin, F-31400 Toulouse, France
[37]Université Grenoble Alpes, CNRS, IPAG, 38000, Grenoble, France
[38]Centro de Astrofísica da Universidade do Porto, Rua das Estrelas, PT4150-762 Porto, Portugal
[39]Centro de Astrobiología (CSIC-INTA), Carretera de Ajalvir km 4, 28850 Torrejón de Ardoz, Madrid, Spain
[40]European Southern Observatory, Karl-Schwarzschild-Straße 2, 85748 Garching bei München, Germany
[41]Physikalisches Institut, University of Bern, Gesellschaftsstrasse 6, CH-3012 Bern, Switzerland

**HIRES will be the high-resolution spectrograph of the European Extremely Large Telescope at optical and near-infrared wavelengths. It consists of three fibre-fed spectrographs providing a wavelength coverage of 0.4-1.8 μm (goal 0.35-1.8 μm) at a spectral resolution of ~100,000. The fibre-feeding allows HIRES to have several, interchangeable observing modes including a SCAO module and a small diffraction-limited IFU in the NIR. Therefore, it will be able to operate both in seeing- and diffraction-limited modes.**

**ELT-HIRES has a wide range of science cases spanning nearly all areas of research in astrophysics and even fundamental physics. Some of the top science cases will be the detection of biosignatures from exoplanet atmospheres, finding the fingerprints of the first generation of stars (PopIII), tests on the stability of Nature's fundamental couplings, and the direct detection of the cosmic acceleration.**

**The HIRES consortium is composed of more than 30 institutes from 14 countries, forming a team of more than 200 scientists and engineers.**

## Introduction

At first light in 2025, the European Extremely Large Telescope (ELT) will be the largest ground-based telescope at visible and infrared wavelengths. The flagship science cases supporting the successful ELT construction proposal were the detection of life signatures in Earth-like exoplanets and the direct detection of the cosmic expansion re-acceleration and it is no coincidence that both science cases require observations with a high-resolution spectrograph.

Over the past few decades high-resolution spectroscopy has been a truly interdisciplinary tool, which has enabled some of the most extraordinary discoveries spanning all fields of Astrophysics, from Planetary Sciences to Cosmology. Astronomical high-resolution spectrometers have allowed scientists to go beyond the classical domain of astrophysics and to address some of the fundamental questions of Physics. In the wide-ranging areas of research exploiting high-resolution spectroscopy, ESO has a long and successful tradition, thanks to the exquisite suite of medium-high resolution spectrographs offered to the community of Member States. UVES, FLAMES, CRIRES, Xshooter and HARPS have enabled European teams to lead in many areas of research. ESPRESSO, which is now joining this suite of very successful high-resolution spectrographs, is fulfilling the promise of truly revolutionising some of these research areas. The scientific interest and high productivity of high-resolution spectroscopy is reflected by the fact that more than 30% of ESO publications can be attributed to its high-resolution spectrographs.

However, it is becoming increasingly clear that, in most areas of research, high-resolution spectroscopy has reached the “photon-starved” regime at 8-10m class telescopes. Despite major progress on the instrumentation front, further major advances in these fields desperately require a larger photon collecting area. Due to its inherently “photon-starved” nature, amongst the various astronomical observing techniques, high-resolution spectroscopy requires the collecting area of Extremely Large Telescopes.

When defining the ELT instrumentation, ESO commissioned two phase-A studies for high-resolution spectrographs, CODEX (Pasquini et al. 2010) and SIMPLE (Origlia et al. 2010) which were started in 2007 and completed in 2010. These studies demonstrated the importance of optical and near-IR high-resolution spectroscopy at the ELT and ESO thus decided to include a High-REsolution Spectrograph (HIRES) in the ELT instrumentation roadmap. Soon after conclusion of the respective phase A studies the CODEX and SIMPLE consortia realized the great scientific importance of covering the optical and near-infrared spectral ranges simultaneously. This marked the birth of the HIRES that started developing the concept of an X-Shooter-like spectrograph, but with higher resolution, capable of providing R ~100.000 over the full optical and near-infrared wavelengths range. Following a community workshop in September 2012 the HIRES Initiative has prepared a White Paper summarizing a wide range of science cases proposed by the community (Maiolino et al. 2013) and also prepared a Blue Book with a preliminary technical instrument concept.

With the start of construction of the ELT, the HIRES Initiative has decided to organize itself as the HIRES Consortium and has recruited additional institutes, which expressed their interest in HIRES. The consortium, strongly motivated by the unprecedented scientific achievements that the combination of such an instrument with the ELT will enable, was commissioned to perform a Phase A study by ESO. The Phase A study started in March 2016 and successfully concluded in May 2018. Following conclusion of the Phase A study, new Institutes from USA and Canada joined the HIRES consortium in the effort of building HIRES.

The HIRES Consortium[a] is now composed of institutes from Brazil, Canada, Chile, Denmark, France, Germany, Italy, Poland, Portugal, Spain, Sweden, Switzerland, United Kingdom and USA. The Italian National Institute for Astrophysics (INAF) is the lead technical Institute. See Marconi et al. (2018) for more details on the Consortium Structure and Organization.

# Science Goals

During the phase-A study, the HIRES Science Advisory Team (SAT), chaired by the Project Scientist, has defined the science priorities for HIRES and determined the corresponding Top-Level Requirements. These science cases, briefly described below, have been then prioritized in order to define the instrument baseline design. Many other science cases are possible with HIRES, but they will not be mentioned here (see, e.g., the community white paper, Maiolino et al. (2013), and Marconi et al. (2018) for a description of the prioritization process) where we focus on few representative science goals.

## Exoplanets and protoplanetary disks

The study of exoplanet atmospheres for a wide range of planetary objects, from gas giants to rocky planets, and from hot to temperate planets, is a primary objective in the field for the next decade. In particular, the detection of components such as molecular oxygen, water and methane in Earth- or super-Earth sized planets is considered to be truly transformational, as they may be regarded as signature of habitability or even signatures of life. Simulations of HIRES observations have been performed in Snellen et al. (2013, 2015) and Hawker & Parry (2019).
HIRES will be able to probe the atmospheres in transmission during the transit of an exoplanet in front of its host star. As an example, it will be possible to detect $CO_2$ absorption in Trappist-1 b with a S/N of 6 in 4 transits of the planet, while $O_2$ absorption at 0.75 μm can be detected in only 25 transits of the planet, i.e. less than 30 hours of observations. HIRES will also be able to directly probe exoplanets, by spatially resolving them from their host star, focussing on their reflected star light and taking advantage of the angular resolution of the ELT with AO-assisted observations. For example, it will be possible to detect the Proxima-Cen b planet in 4 nights of integration with a S/N of 8 with a relatively simple system of single conjugate adaptive optics (SCAO), similar to that used by other ELT first-light instruments. Figure 1, left, shows that HIRES will be able to detect O2 from a Proxima-b like exoplanet in 70 h of integration.

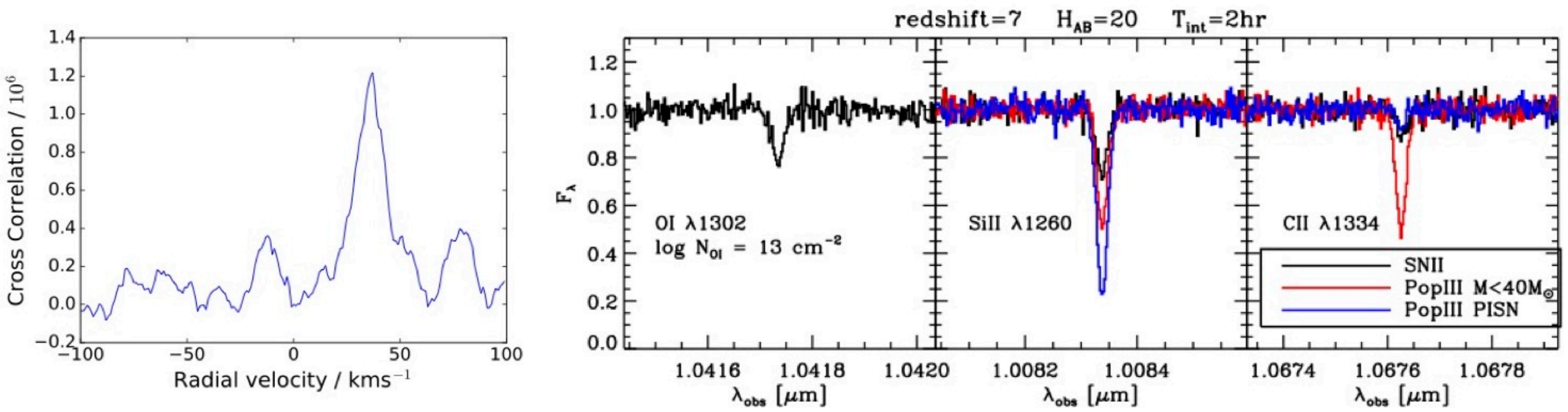

**Figure 1.** *HIRES science highlights. Left: Cross Correlation Signal indicating the clear detection of $O_2$ in as Proxima-b like exoplanet in 70h of total integration (adapted from Fig. 4 of Hawker & Parry 2019). Right: Observations of a z=9 quasar with $H_{AB}$=22 and a total integration time of 10h showing HIRES capability of distinguishing IGM enrichment by normal SNII supernovae or by low mass and pair instability Supernovae from Pop III Stars (simulations by the HIRES Science Advisory Team).*

Protoplanetary disks are a natural outcome of angular momentum conservation in star formation and are ubiquitous around young, forming stars. HIRES will be able to determine the properties of the gas in the inner star-disk region, where different competing mechanisms of disk gas dispersal are at play. This will constrain on one side on the mechanisms through which the forming star acquires mass and removes the angular momentum, and on the other side the initial condition for planets formation.

## Stars and Stellar Populations

The vast light-collecting power of the ELT will enable detailed high-resolution spectroscopy of individual stars, and in particular very faint red dwarfs and distant red giants in nearby galaxies, for which HIRES will be able to provide tight constraints for the atmospheric parameters. These constraints will be extremely important to characterize the stellar hosts of exoplanets.
HIRES will also expand our horizon by measuring the heavy-elements abundances of the most primitive stars (low mass, low metallicity) in our galaxy and its satellites helping us to understand what is the lowest metallicity for which gas can collapse to form low-mass stars, and what are the nature and yields of the very first generation of stars and their supernovae.
Last, but not least, the combination of very high spectral resolving power and diffraction-limited angular resolution makes the ELT a unique resource for deepening our understanding of the physics of stellar atmospheres and nucleosynthesis processes, by allowing to spectroscopically resolve the effects of surface convection and to measure isotopic abundances of atomic species.

## Galaxy Formation and evolution and the intergalactic medium

The detection of the first generation of stars (so called Pop III stars) and the observational characterization of their properties is one of the main objectives of extragalactic astrophysics. Proto-galaxies hosting Pop III stars are expected to be too faint for direct detection, even with JWST. However, the signature of Pop III stars can be detected through their nucleosynthetic yields which can be potentially observed in the abundance patterns of very metal-poor absorption systems in the high-resolution, wide-range spectra of bright high-redshift sources provided by HIRES in the NIR (Figure 1, right).

The direct detection and characterization of the beginning of the reionization epoch is another very important goal in the study of galaxy formation. This process is believed to have been dominated by ultraviolet photons from the first generations of galaxies, most of which are too faint to be observed directly even with JWST. By targeting bright quasars at high redshift as background continuum sources, HIRES will be able to study both transmission features in the Lyman-α forest and metal absorption lines associated with these reionization-epoch sources, constraining the patchiness of the reionization process, the properties of the ultraviolet background radiation and the chemical enrichment of the IGM in this epoch.

## Cosmology and Fundamental Physics

The observational evidence for the acceleration of the expansion of the universe and the tensions that have been highlighted by different cosmological probes have shown that our canonical theories of cosmology of fundamental physics may be incomplete (and possibly incorrect), and that there might be unknown physics yet to be discovered. HIRES will allow to search for, identify and ultimately characterize any new physics through several different but fundamentally inter-related observations which will enable a unique set of tests of the current cosmological paradigm.

HIRES will be able to constrain the variation of fundamental physical constants like the fine-structure constant $\alpha$ and proton-electron mass ratio $\mu$ with the advantage, compared to laboratory measurements, of exploring variations over 12 Gyr timescales and 15 Gpc spatial scales. A detection of varying fundamental constants would be revolutionary: it would automatically prove that the Einstein Equivalence Principle is violated (i.e. gravity is not purely geometry), and that there is a fifth force.

HIRES will enable a test of the CMB temperature-redshift relation, $T(z) = T_0 (1 + z)$, which is a robust prediction of standard cosmology but that must be directly verified by measurements. A departure from this relation can in turn reveal a violation of the hypothesis of local position invariance (and thus of the equivalence principle) or that the number of photons is not conserved. HIRES measurement will greatly improve on the existing constraints on T(z) compared to existing data.

The redshifts of cosmologically distant objects drift slowly with time (the so-called Sandage effect). A redshift drift measurement is fundamentally different from all other cosmological observations and can provide a direct detection of cosmic reacceleration, thus undoubtedly confirming cosmic acceleration, the existence of dark energy and potentially provide evidence for new physics. HIRES will be capable of detecting the redshift drift in the Lyα forests of the brightest currently known QSOs (~6 cm/s/decade at $z = 4$ for a Planck-like standard cosmology). The ELT may thus become the first facility ever to watch the Universe change in "real time".

## Science Priorities

These are just a few of the many science cases that can be addressed, a collection of many of these can be found in the community white paper (Maiolino et al. 2013). However, in order to define the instrument baseline design a prioritization of the science cases was performed by the HIRES Science Advisory Team following criteria of scientific impact (transformational versus incremental), feasibility and competitiveness.
Then, if the TLR's of the top priority science cases were enabling other science cases, the latter were not considered any further in the subsequent prioritization, as considered accomplished together with the top priority science cases.
The top science priorities and associated requirements are listed below:

1) **Exoplanet atmospheres in transmission,** requiring a spectral resolution of at least 100,000, a wavelength coverage of at least 0.50-1.80 μm and a wavelength calibration accuracy of 1 m/s. The implementation of the above TLRs would automatically enable the following science cases:
    - **reionization of the universe,**
    - **the characterization of cool stars,**
    - **the detection and investigation of near pristine gas,**
    - **the study of Extragalactic transients.**
2) **Variation of the Fundamental Constants of Physics,** requiring an extension to 0.37 μm in addition to the TLRs of priority 1. These extension towards the blue would also automatically enable to investigate:
    - **the cosmic variation of the CMB temperature,**
    - **the determination of the deuterium abundance,**
    - **the investigation and characterization of primitive stars.**

    At λ < 0.40 μ m the throughput of the ELT is expected to be low as a consequence of the planned coating. However, even in the range 0.37-0.40 μm the system is expected to outperform ESPRESSO at the VLT, and new coating is under study and may be available a few years after first light.
3) **Detection of exoplanet atmospheres in reflection,** requiring, on top of the TLRs of priority 1, the addition of an Adaptive Optics (SCAO) system and an Integral Field Unit. Reflected-light spectra allow tracing atmospheric emission from lower altitudes on the dayside of the exoplanet. These additional TLRs would automatically enable also the following cases:
    - **Planet formation in protoplanetary disks,**
    - **Characterization of stellar atmospheres,**
    - **Search of low mass Black Holes.**
4) **Sandage test.** Its additional TLRs, are a wavelength range of 0.40-0.67 μm and a stability of 2 cm/s, enabling also:
    - **the mass determination of Earth-like exoplanets**
    - **radial velocity searches and mass determinations for exoplanets around M-dwarf stars.**

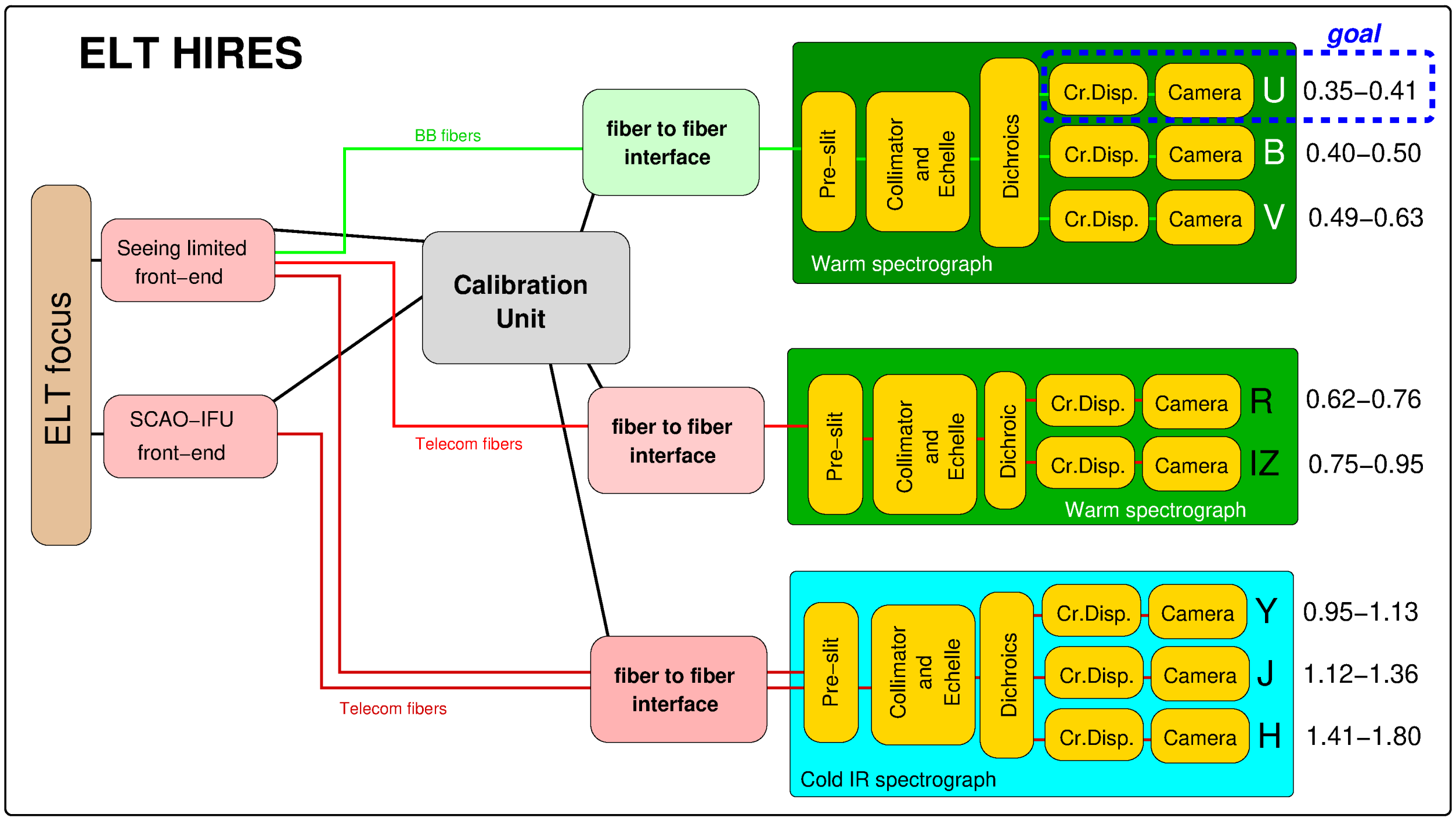

**Figure 2.** *HIRES architectural design, outlining the instrument subsystems: Front End (seeing-limited and AO assisted with SCAO unit), Fibre Link, Calibration Unit, VIS-Blue, VIS-Red and NIR (cold spectrograph).*

## Instrument concept

Following phase A and further studies before the start of construction, the HIRES baseline design is that of a modular instrument consisting of three fibre-fed cross dispersed echelle spectrographs VIS-BLUE (UBV), VIS-RED (RIZ) and NIR (YJH), providing a simultaneous spectral range of 0.4-1.8 µm (goal 0.35-1.8 µm) at a resolution of 100,000. The fibre feeding allows several, interchangeable, observing modes ensuring maximization of either accuracy, throughput or spatially resolved information. Together with the SCAO module, the proposed baseline design capable of fulfilling the requirements of the 4 top science cases.

The baseline design is summarized below but several alternatives have been evaluated during the Phase A study. Also, several add-ons made possible by the modular nature of the instrument have been considered (e.g. a polarimetric module in the intermediate focus, or a wavelength extension out to the K-band 2.0-2.4 µm).
The overall concept is summarized in Figure 2: in the Front End the light from the telescope is split, via dichroics, into 3 wavelength channels. Each wavelength channel interfaces with several fibre bundles that feed the corresponding spectrograph module. Each fibre-bundle corresponds to an observing mode and all together they constitute the Fibre Link. All spectrographs, VIS-BLUE, VIS-RED and NIR, have a fixed configuration, i.e. no moving parts, allowing to fulfil the requirements on stability. They include a series of parallel entrance slits consisting of linear micro-lens arrays each glued to the fibre bundles. The split in wavelengths between the spectrographs is influenced, among other parameters by the optical throughput of the different types of fibres available on the market; therefore, the different modules can be positioned at different distances from the focal plane of the telescope.
The whole instrument should be placed on the Nasmyth platform, if enough volume and mass is available. If necessary, the fibre feeding allows the VIS-RED and NIR modules to be placed in the Coudé Room, which can also host the Calibration Unit.

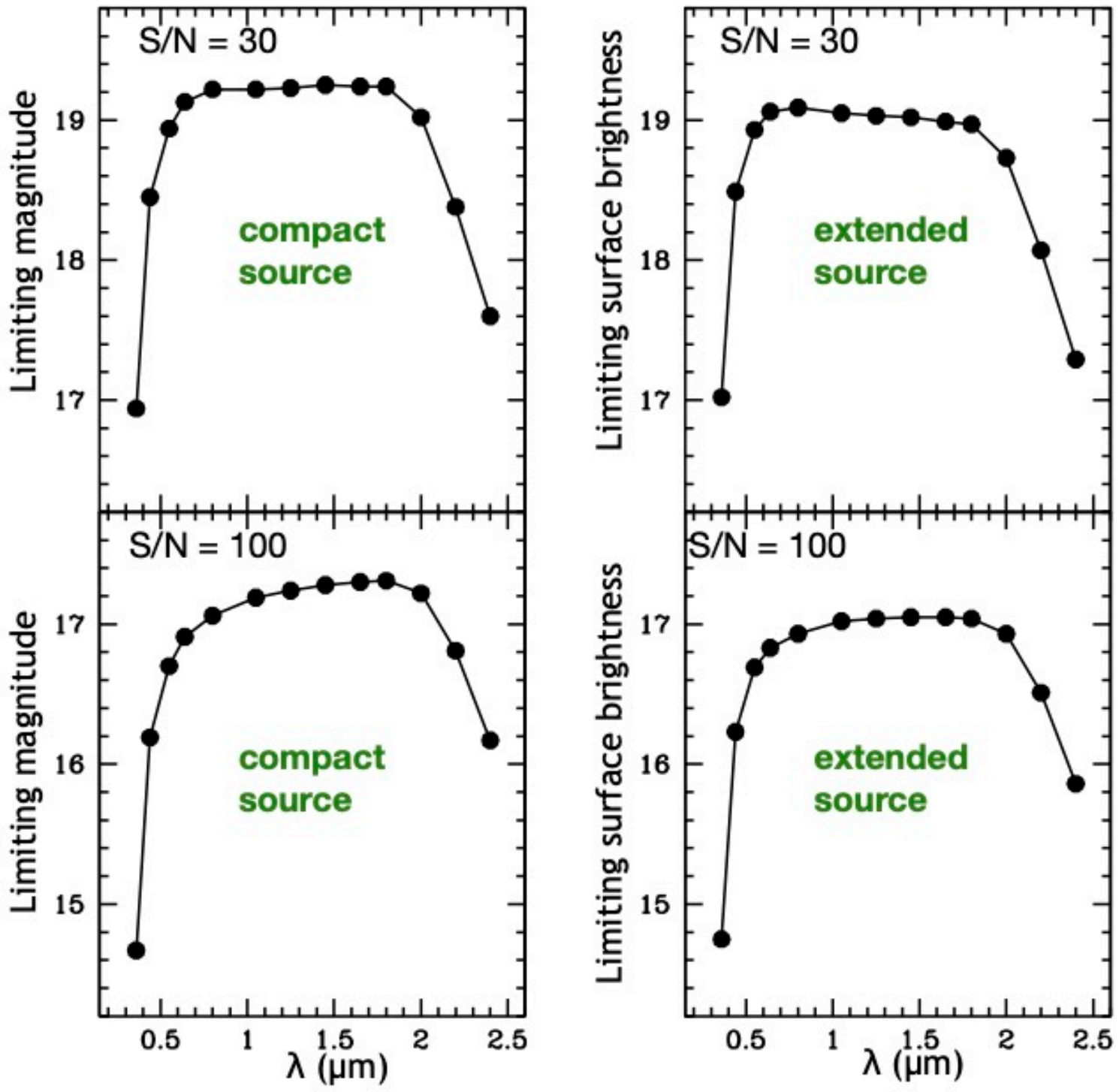

**Figure 3.** *HIRES limiting magnitudes obtained from the ETC for different S/N ratios (30 – top and 100 – bottom), compact and extended sources (left and right). Observations are in seeing-limited mode with R = 100,000 a total exposure time of 1800s.*

# Performance

The Exposure Time Calculator, regularly updated to take into account modifications in the design, is maintained by INAF-Arcetri and can be run at the http://hires.inaf.it/etc.html web link. This ETC can compute the limiting magnitude achievable at a given wavelength, in a given exposure time and at a given signal to noise ratio or it can compute the signal to noise ratio achievable at a given wavelength, in a given exposure time and at a given magnitude. HIRES expected performances computed with the ETC are summarized in Figure 3.

## Conclusions

The HIRES baseline design is that of three ultra-stable and modular fibre-fed cross dispersed echelle spectrographs providing a simultaneous spectral coverage of 0.4-1.8 µm (goal 0.35-1.8 µm) at a resolution of 100,000 with several, interchangeable, observing modes ensuring maximization of either accuracy, throughput or spatially resolved information. Overall, the studies conducted so far have shown that the HIRES baseline design can address the 4 top priority science cases, being able to provide ground-breaking science results with no obvious technical showstoppers.

The construction of HIRES includes the majority of the institutes in ESO member states with expertise in high resolution spectroscopy and will require an estimated 30 MEUR in hardware (excluding contingencies) and about 500 FTEs. Contingencies are expected to be low (5-10%) because the proposed baseline design is based on proven technical solutions and can benefit on heritage from HARPS and ESPRESSO and other previous high-resolution spectrographs, e.g. PEPSI at the 11.8m LBT, SPIRou and CARMENES. The construction will last about 8-10 years. Therefore, with Phase B starting in 2021, HIRES could be at the telescope as early as 2030.

Overall, HIRES is an instrument capable of addressing ground-breaking science cases while being almost (telescope) pupil independent, as it can operate both in seeing and diffraction limited modes; the modularity ensures flexibility during construction and the possibility to quickly adapt to new development in the technical as well as science landscape.

### Acknowledgements

The Italian effort for HIRES is supported by the Italian National Institute for Astrophysics (INAF).
HIRES work in the UK is supported by the Science and Technology Facilities Council (STFC) at the UKATC, University of Cambridge (grants ST/S001387/1 and ST/N002873/1) and Heriot Watt University (grant ST/S001328/1), as part of the UK E-ELT Programme.
We acknowledge financial support from the Spanish Ministry of Science and Innovation (MICINN) projects AYA2017-86389-P, RYC-2013-14875, PGC2018-098153-B-C31, PID2019-109522GB-C51/52.
The German efforts for HIRES are funded by the Federal Ministry for Education and Research (BMBF). KGS thanks the BMBF-Verbundforschung for support through grants 05A17BAB and 05A2020.
This work was supported by FCT - Fundação para a Ciência e a Tecnologia through national funds and by FEDER through COMPETE2020 - Programa Operacional Competitividade e Internacionalização by these grants: UID/FIS/04434/2019; UIDB/04434/2020; UIDP/04434/2020; PTDC/FIS-AST/32113/2017 & POCI-01-0145-FEDER-032113; PTDC/FIS-AST/28953/2017 & POCI-01-0145-FEDER-028953; PTDC/FIS-AST/28987/2017 & POCI-01-0145-FEDER-028987.

Research activities of the observational astronomy board at the Federal University of Rio Grande do Norte are supported by continuous grants from the Brazilian funding agencies CNPq, FAPERN, and INCT-INEspaço. This study was financed in part by the Coordenação de Aperfeiçoamento de Pessoal de Nível Superior - Brasil (CAPES) - Finance Code 001.

## Links

[1] Instrument Web Page http://hires.inaf.it
[2] Exposure time calculator http://hires.inaf.it/etc.html

## Notes

[a] Partners of the HIRES Consortium (CI = Coordinating Institute within a country)
**Brazil:** Núcleo de Astronomia Observacional, Universidade Federal do Rio Grande do Norte (CI); Instituto Mauá de Tecnologia. **Canada:** Institut de Recherche sur les Exoplanètes and Observatoire du Mont-Mégantic, département de physique, Université de Montréal. **Chile:** Pontificia Universidad Catolica de Chile (CI); Centre of Astro Engineering, Universidad de Chile; Department of Astronomy, Universidad de Concepcion; Center of Astronomical Instrumentation, Universidad de Antofagasta. **Denmark:** Niels Bohr Institute, University of Copenhagen (CI); Department of Physics and Astronomy, Aarhus University. **France:** Laboratoire d’Astrophysique de Marseille, CNRS, CNES, AMU (CI); Institut de Planétologie et d’Astrophysique de Grenoble, Université Grenoble Alpes; Laboratoire Lagrange, Observatoire de la Côte d’Azur; Observatoire de Haute Provence, CNRS, AMU, Institut Pythéas, Institut de Recherche en Astrophysique et Planetologie, Observatoire Midi-Pyrénées; Laboratoire Univers et Particules, Université de Montpellier. **Germany:** Leibniz-Institut für Astrophysik Potsdam (CI); Institut für Astrophysik, Universität Göttingen; Zentrum für Astronomie Heidelberg, Landessternwarte; Thüringer Landesternwarte Tautenburg; Hamburger Sternwarte, Universität Hamburg. **Italy:** INAF, Istituto Nazionale di Astrofisica (Lead Technical Institute). **Poland:** Faculty of Physics, Astronomy and Applied Informatics, Nicolaus Copernicus University in Torun. **Portugal:** Instituto de Astrofísica e Ciências do Espaço (IA) at Centro de Investigaço em Astronomia/Astrofísica da Universidade do Porto (CI), Instituto de Astrofísica e Ciências do Espaço at Faculdade de Ciências da Universidade de Lisboa. **Spain:** Instituto de Astrofísica de Canarias (CI); Instituto de Astrofísica de Andalucía-CSIC; Centro de Astrobiología **Sweden:** Dept. of Physics and Astronomy, Uppsala University. **Switzerland:** Département d'Astronomie, Observatoire de Sauverny, Université de Genève (CI); Universität Bern, Physikalische Institut. **United Kingdom:** Science and Technology Facilities Council (CI); Cavendish Laboratory & Institute of Astronomy, University of Cambridge; UK Astronomy Technology Centre; Institute of Photonics and Quantum Sciences, Heriot-Watt University. **USA:** Department of Astronomy, University of Michigan.